\documentclass[10pt,a4paper,twoside]{article}

\usepackage{vmargin,fancyheadings,multicol,multirow,ifthen,cite,
    epsfig,wrapfig,calc,dcolumn,apalike,graphicx,
    setspace, 
    boxedminipage,rotating,
    textcomp, 
        booktabs, 
    picinpar,longtable}

\usepackage{wgn2} 

\begin{document}
\SetPaperBodyFont   



\setcounter{page}{1}

\begin{WGNpaper}{twocol}{

\title{The Perseid and Geminid meteor shower activity over Hungary in 2019-2023}

\author{
L\'ivia Deme, Kriszti\'an S\'arneczky, Antal Igaz, Bal\'azs Cs\'ak, N\'andor Opitz, N\'ora Egei, J\'ozsef Vink\'o
\thanks{HUN-REN CSFK Konkoly Observatory, Konkoly Thege \'ut 15-17, Budapest, 1121 Hungary \\
        Email: {\tt vinko@konkoly.hu}}
}

\abstract{We present statistical analysis of video meteor observations 
for the Perseid and Geminid showers taken with two camera systems 
operating in Hungary from the end of 2019 through 2023. Zenithal hourly
rates (ZHR) and meteor fluxes, determined by MetRec-based analog video cameras
{\tt HUKON}, {\tt HUPIS} and {\tt HUHOD}, are inferred and compared with
detections of slow fireballs measured at the same sites by a system consisting of automated DSLR cameras (the KoMON system). }
\datereceived{Received 2024 May 31}
}
\section{Introduction}
    \thankstext{HUN-REN CSFK Konkoly Observatory, Konkoly Thege \'ut 15-17, Budapest, 1121 Hungary \\
        Email: {\tt vinko@konkoly.hu}}
        
    \setcounter{footnote}{0} 

This paper is the second part of a series presenting the recent meteor activity over Hungary, based on video meteor data taken with different automated camera systems from 2019 December through 2023 November. In the first part (\cite{deme23}) we summarized the details of the three camera systems owned and operated by our group: a set of MetRec-based analog video cameras, {\tt HUKON}, {\tt HUPIS} and {\tt HUHOD}, the Konkoly Meteor Observatory Network (KoMON) consisting of 18 autonomous, computer- controlled DSLR-cameras at 5 different locations, and 5 nodes of the AllSky7 system distributed within Hungary. For the basic parameters and some technical details on these camera systems see \cite{deme23}. 

In \cite{deme23} we also presented a statistical analysis of sporadic meteors based on the data taken by these camera systems between 2019 and 2023. 

In the present paper we explore the properties of two of the richest meteor showers, the Perseids (PER) and the Geminids (GEM), during the same years. The basic parameters of these two showers are summarized in Table~\ref{tab:basic}. Note that the given radiant coordinates represent the values near the peak activity, but the actual physical structure of both streams are more complex. The PER shower, for example, consists of at least three components (\cite{rendtel97}): a relatively weak background component that is longer than a month, a high-activity peak that lasts for $\sim2$ days and an outburst component lasting only for a few hours. On the contrary, the GEM shower extends for only $\sim 10$ days and the peak lasts for less than a day (\cite{rendtel04}). Both the position and the $ZHR$ of the GEM peak are quite stable, which is consistent with the age estimate of $\sim 6000$ years for the GEM stream. 

We used data from our three MetRec cameras, as well as from the KoMON cameras at two sites, Budapest and Piszkesteto, which are at the same location as {\tt HUKON} and {\tt HUPIS}. 

\section{Methods}

First, the population index ($r$) for the two showers were determined from the data taken by the {\tt HUKON} camera (see \cite{deme23} for details) using the following definition (e.g. \cite{br96})
\begin{equation}
    r ~=~ \frac{N(m+1)}{N(m)},
    \label{eq:popindex}
\end{equation}
where $N(m)$ is the number of meteors having peak brightness $m_{peak} \le m$. From the population index one can estimate the mass index ($s$) of the shower following \cite{ceplecha98} 
\begin{equation}
    s ~=~ 1 + 2.5 \log_{10} r.
    \label{eq:massindex}
\end{equation}
The mass index is related to the mass distribution of the meteoroids in a particular shower, which is found to be a power-law with index $-s$ (e.g. \cite{vida22}).

Having the population index determined, the number of observed meteors detected by a camera with limiting magnitude $m_{lim}$ can be extrapolated to $m = 6.5$ mag as 
\begin{equation}
    N_{met} ~=~ n_{obs} \cdot r^{6.5 - m_{lim}}.
    \label{eq:limmag}
\end{equation}

\begin{table}[]
    \centering
    \begin{tabular}{lcc}
         Parameter & PER & GEM  \\
        \hline 
         Parent body & 109P/Swift-Tuttle & (3200) Phaethon \\
         Radiant R.A. & $3^h 13^m$ &  $7^h 28^m$ \\
         Radiant Dec. & $+58^o$ &  $+32^o$ \\
         Peak date & Aug 13 & Dec 12 \\
         Mean velocity & 58.8 km/s & 35.0 km/s\\
         Peak ZHR & $\sim 100$ & $\sim 120$ \\         
         \\
    \end{tabular}
    \caption{Basic parameters of the PER and GEM meteor showers. The given radiant coordinates refer to the period near the peak/maximum date.}
    \label{tab:basic}
\end{table}

Finally, the Zenithal Hourly Rate for the meteor shower can be expressed as
\begin{equation}
    ZHR ~=~ \frac{N_{met}}{t_{eff}} \cdot F \cdot 
    \frac{1}{sin^{\gamma} h_R},
    \label{eq:zhr1}
\end{equation}
where $F$ is the geometric correction factor that scales the field-of-view (FoV) of a given camera to the average FoV of a visual observer.  For the latter we adopted 1/3 of the area of the visible sky hemisphere, while for the former we used the camera FoV values as given in Table~1 in \cite{deme23}. In Eq.\ref{eq:zhr1} $t_{eff}$ is the effective observing time during the night (corrected for interruptions due to clouds), $h_R$ is the elevation of the shower radiant and $\gamma \geq 1$ is the zenith exponent. In this study we assumed $\gamma = 1$, thus, our $ZHR$ estimates are actually lower limits. 

Since $h_R$ is varying during the night, in principle, Eq.\ref{eq:zhr1} should be computed during a sufficiently short time interval when the change in $h_R$ is negligible. However, such an approach requires a lot of data in order to get a sufficiently rich statistical sample, which is feasible only during multi-station observing campaigns, or in multi-station camera networks. Since we used single-station camera data for the present study, we took a different approach. We computed the $h_R$ values at the moment of each individual meteor detection, and summed up all the events to get the $ZHR$ value as follows:
\begin{equation}
    ZHR ~=~ \frac{F \cdot r^{6.5 - m_{lim}}}{t_{eff}} 
    \sum_{i=1}^{n_{obs}}{ \frac{1}{\sin^{\gamma} h_R(i)} },
    \label{eq:zhr2}
\end{equation}
where $h_R(i)$ is the  elevation of the radiant at the moment of the detection of the $i$th meteor. In order to avoid assigning too high weight to meteors when the radiant was at low altitude, we set $h_R = 20$ deg as a lower limit during the calculations. 

We also estimated the meteoroid fluxes from the formula given by \cite{kr90}:
\begin{equation}
    F_{6.5} ~=~ \frac{ZHR}{37200} \cdot (13.1 \cdot r - 16.5) \cdot (r-1.3)^{0.748},
\end{equation}
where $F_{6.5}$ is the meteoroid flux to the magnitude limit of 6.5 in units of meteoroids~km$^{-2}$~h$^{-1}$. 

Finally, the meteoroid limiting mass $M_{lim}$ corresponding to a limiting magnitude $m_{lim}$ mag was computed as 
\begin{equation}
    \log_{10} (M_{lim}) ~=~ \frac{-8.75 \log_{10}(v_0) - m_{lim} + 11.59}{2.25},
    \label{masslim}
\end{equation}
following \cite{vida22}, where $v_0$ is the velocity of the meteoroid. Since we have no velocity information on individual meteoroids, we applied the mean shower velocities listed in Table~\ref{tab:basic}. 

\section{Results}

In this section we describe our results.

\subsection{Perseids}

\begin{figure*}
    \centering
    \includegraphics[width=17.5cm]{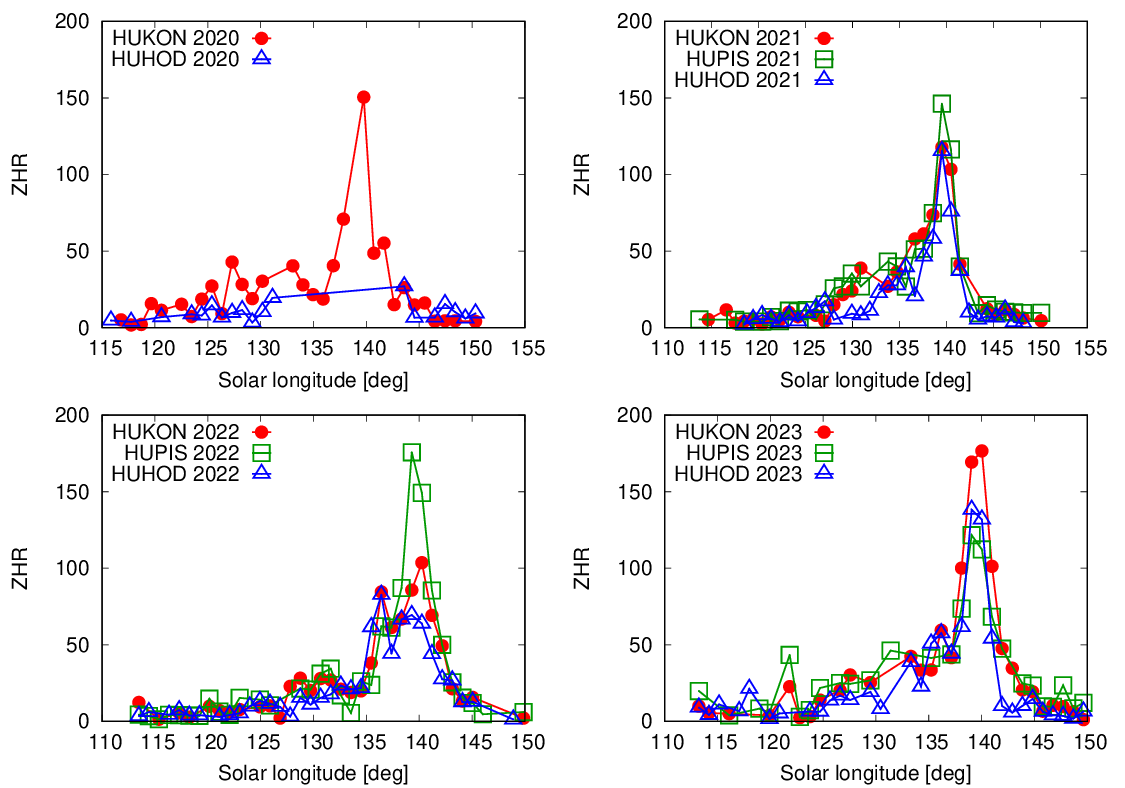}
    \caption{The ZHR of the PER shower measured by the MetRec-based cameras.}
    \label{fig:perzhr}
\end{figure*}

Table~\ref{tab:popindex_per} shows the population indices for the PER shower derived from the {\tt HUKON} camera data.  These are computed from the whole dataset obtained each year within the solar longitude interval of $\lambda_\odot = 113$ deg and $150$ deg. Since our data are dominated by meteors near the activity peak (from $\lambda_\odot = 135$ deg to $145$ deg), these $r$ values represent the mean population index around the peak. It is known that the population index for the PER shower varies along the crossing of the stream (e.g. \cite{rendtel97}), thus, our values in Table~\ref{tab:popindex_per} can be considered only as a weighted average for each year. Nevertheless, these values are in good agreement with other previously published data (see e.g. Table~1 in  \cite{vida22}). Hereafter we adopt $r = 2.0$ as the mean value for the PER meteors, which corresponds to a mass index of $s = 1.75$. 

\begin{table}[]
    \centering
    \begin{tabular}{ccc}
     Year & $r$ & uncertainty \\
    \hline
     2020 & 1.99 & 0.75 \\
     2021 & 2.09 & 0.62 \\
     2022 & 1.89 & 0.55 \\
     2023 & 2.13 & 0.65 \\
     All  & 2.04 & 0.63 \\
     \\
    \end{tabular}
    \caption{Measured mean population indices for the Perseids detected by {\tt HUKON}}
    \label{tab:popindex_per}
\end{table}

Fig.~\ref{fig:perzhr} displays the measured nightly $ZHR$ values (calculated via Eq.(\ref{eq:zhr2})) as a function of solar longitude for each year. Bad weather prevented the collection of more data at the sites of {\tt HUPIS} and {\tt HUHOD} in 2020. 

In Table~\ref{tab:zhr_per} we list the solar longitudes at maximum activity, the maximum $ZHR$ values and $F_{6.5}$ fluxes (in $10^{-3}$~km$^{-2}$~h$^{-1}$ units). These values were inferred via fitting Gaussian functions to the measured nightly $ZHR$ values near the maximum for each camera. 
The limiting mass, according to Eq.(\ref{masslim}), is $M_{lim} = 2.14 \times 10^{-5}$ g at the limiting magnitude of $+6.5$ mag. 

\begin{table}
\centering
\begin{tabular}{lccc}
Year & $\lambda_\odot$(max) & $ZHR$(max) & $F_{6.5}$(max) \\
     & (deg) &  & ($10^{-3}$~km$^{-2}$~h$^{-1}$) \\ 
\hline
{\tt HUKON} & & & \\
2020 & 139.3 & 155.7 & 31.0 \\
2021 & 139.7 & 125.6 & 25.1 \\
2022 & 140.0 & 116.6 & 23.3 \\
2023 & 139.6 & 180.8 & 36.1 \\
\\
{\tt HUPIS} & & & \\
2021 & 139.7 & 144.9 & 28.9 \\
2022 & 139.7 & 184.8 & 36.9 \\
2023 & 139.5 & 124.1& 24.9 \\
\\
{\tt HUHOD} & & & \\
2021 & 139.6 & 115.6 & 23.1 \\
2022 & 139.3 & 85.9 & 17.1 \\
2023 & 139.4 & 140.1 & 28.0 \\
\\
\end{tabular}
\caption{Measured parameters of the PER shower, assuming $r = 2.0$ ($s=1.75$) and $\gamma=1$.} 
\label{tab:zhr_per}
\end{table}

\cite{vida22} reported a PER outburst in 2021, reaching $ZHR \approx 277 \pm 18$ for a short period of time ($\sim 2.5$ hours) at solar longitude $\lambda_\odot = 141.47$ deg.
This was confirmed by several other authors (e.g. \cite{misk21-2}; \cite{jm21}). 
Unfortunately, since this interesting event was visible only from a few sites in North America, our cameras have not detected it. 

\begin{table*}[]
    \centering
    \begin{tabular}{lccl}
    Year & $\lambda_\odot$(max) & $ZHR$(max) & Reference \\
    \hline
    2020 & 139.3 & 155 & this paper \\
    2020 & 140.4 & 125 & \cite{vida22} \\
    2020 & 140.7 & 90 &  \cite{misk21} \\
    2021 & 139.7 & 145 & this paper \\
    2021 & 140.2 & 135 & \cite{vida22} \\
    2021 & 141.5 & 277 & \cite{vida22} \\
    2021 & 141.5 & 195 & \cite{misk21-2} \\
    2021 & 141.5 & 210 & \cite{jm21} \\
    2022 & 139.7 & 184 & this paper \\
    2022 & 139.9 & 110 & \cite{rendtel23} \\
    2023 & 139.6 & 181 & this paper \\
    \\
    \end{tabular}
    \caption{Maximum $ZHR$ values and the solar longitudes at maximum for the PER meteors. The second maximum from 2021 corresponds to the PER outburst reported by North American observers.}
    \label{tab:per_zhr_max}
\end{table*}

In Table~\ref{tab:per_zhr_max} the maximum $ZHR$ values and the solar longitudes at maximum are collected for the PER shower. Our $ZHR$ values seem to be systematically higher than the literature data, except for the 2021 outburst. However, since the outburst was not captured by our cameras, our slightly higher value in 2021  does not correspond to the increased activity. Instead, this systematic effect could be due to the adopted field-of-view correction factor $F$, which is $\sim 1.5$ for our MetRec cameras. Without this factor, our $ZHR$ estimates would be somewhat lower than the literature data, thus, the real values are somewhere in between these two limits. 

\subsection{Geminids}

\begin{figure*}
    \centering
    \includegraphics[width=17.5cm]{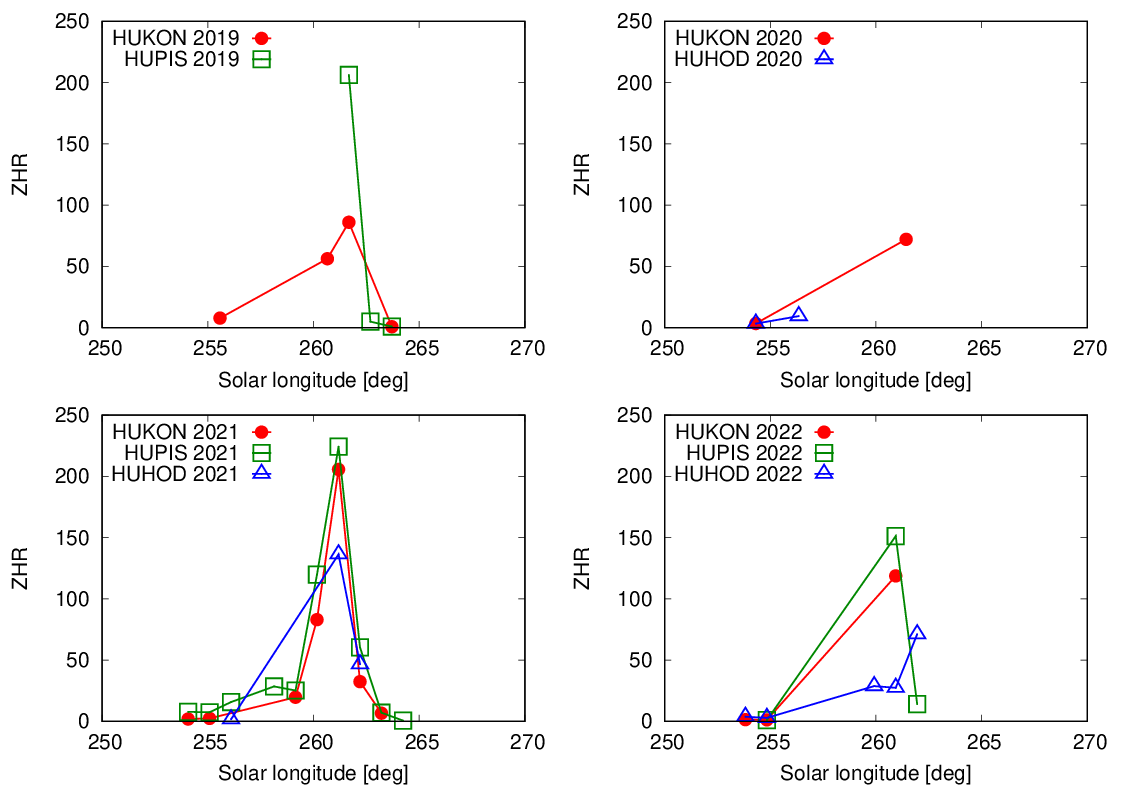}
    \caption{The ZHR of the GEM shower measured by the MetRec-based cameras. }
    \label{fig:gemzhr}
\end{figure*}

\begin{figure*}
    \centering
    \includegraphics[width=17.5cm]{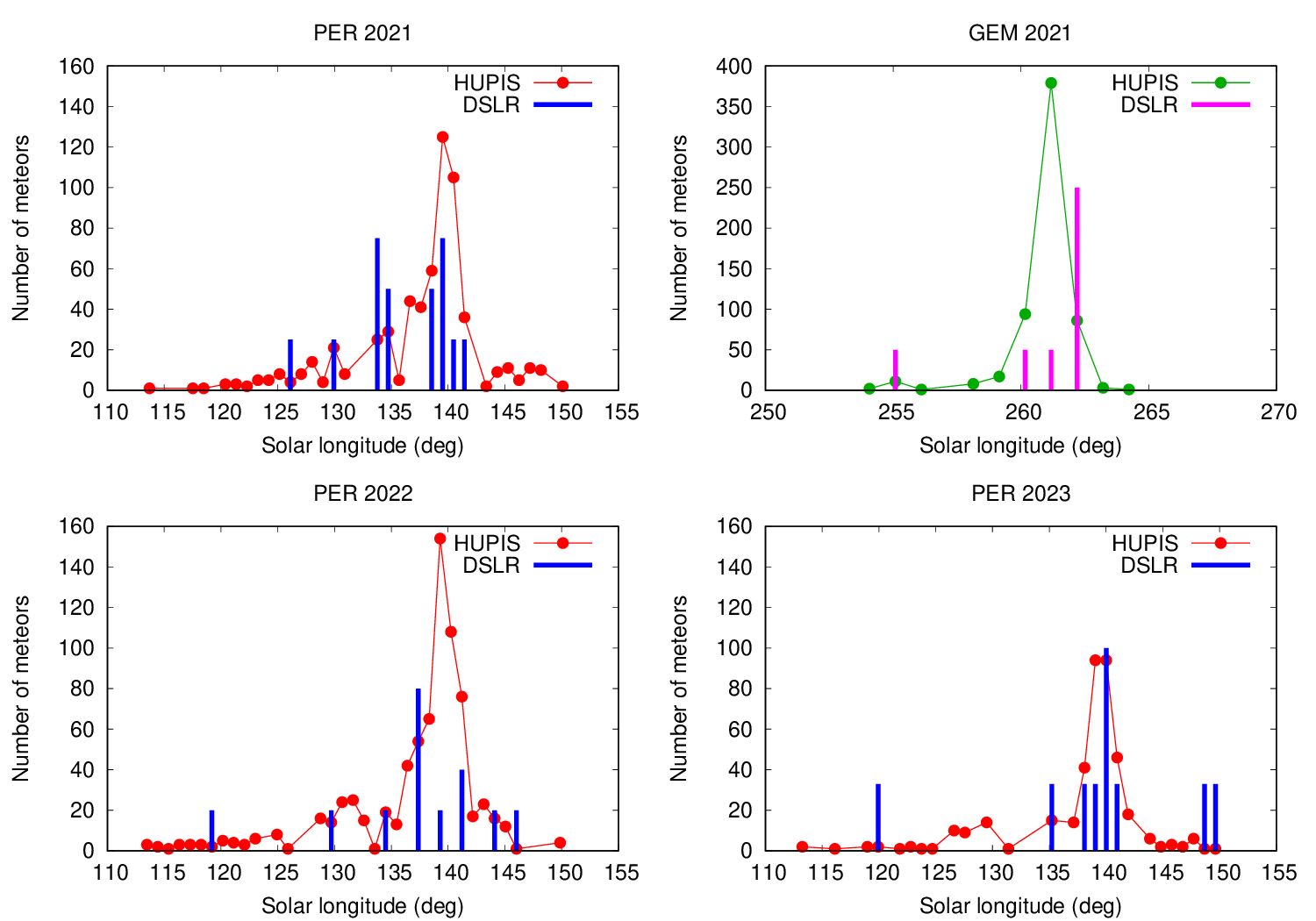}
    \caption{Comparison of meteor numbers detected by MetRec (dots) and KoMON (impulses). The KoMON data are scaled up with a factor of 100 to enhance their visibility.}
    \label{fig:metrec_dslr_comp}
\end{figure*}

Mid-December, the time of peak activity of the GEM shower, is usually affected by bad weather (clouds, fog) in Central Europe. Therefore, our GEM data are much more sparse than the PER data discussed above. 

\begin{table}[]
    \centering
    \begin{tabular}{ccc}
     Year & $r$ & error \\
    \hline
     2019 & 3.86 & 0.93 \\
     2020 & 2.00 & 1.08 \\
     2021 & 2.75 & 0.80 \\
     2022 & 2.45 & 0.51 \\
     All  & 2.67 & 0.74 \\
     \\
    \end{tabular}
    \caption{Measured mean population indices for the Geminids detected by {\tt HUKON}.}
    \label{tab:popindex_gem}
\end{table}

The population indices derived for the GEM data are collected in Table~\ref{tab:popindex_gem}. Again, these are weighted averages computed during solar longitudes from $\lambda_\odot = 254$ deg to 264 deg each year. Low number statistics due to bad weather is reflected by their higher uncertainties. Overall, these indices are higher than some of the literature values ($r \sim 2$, \cite{vida22}), but other measurements are closer to our values (e.g. $r_{mean} \sim 2.3$, \cite{watanabe18}). Since $r$ is varying during the activity, such deviations between the mean values, measured by different observers at different conditions, can be expected.
Because of the higher uncertainties involved in these estimates, we adopted the lower literature value, $r = 2.0$ for the GEM shower for the rest of the analysis.

Fig.~\ref{fig:gemzhr} shows the estimated $ZHR$ profiles for the GEM shower. We were able to collect usable amount of data only in 2021, when we measured the peak activity at 
$\lambda_\odot = 261.0$ deg, and $ZHR = 213$ and $229$ corresponding to $F_{6.5} = 42.6$ and $45.7 \times 10^{-3}$~km$^{-2}$~h$^{-1}$ fluxes for {\tt HUKON} and {\tt HUPIS}, respectively. These fluxes correspond to the limiting mass of $M_{lim} = 1.75 \times 10^{-4}$ g for the GEM meteoroids. Our solar longitude agrees well with the value reported by \cite{vida22} ($\lambda_\odot = 261.98$) for 2021, but our maximum $ZHR$ data are almost twice as high as the one measured by \cite{vida22} ($ZHR_{max} \sim 128$). Since the GEM measurements could be heavily affected by bad weather all around the northern hemisphere, it is not clear what is behind the disagreement. If we disregard the field-of-view correction factor ($F = 1.5$), our $ZHR_{max}$ estimate reduces to $\sim 147$, which is still somewhat higher than the one reported by \cite{vida22}. Note that \cite{watanabe18} also measured a similarly high peak $ZHR \sim 230 \pm 50$ for the GEM shower in 2013, compared to the preceding years.

\subsection{The frequency of slow fireballs during the PER and GEM showers}

The KoMON meteor camera system is intended to capture the brightest slow meteors, many of them appear as fireballs (flashes) (see \cite{deme23} for a more detailed description). Here we collected the data from the KoMON cameras installed at Piszk\'estet\H{o}, which is the same site as that of the {\tt HUPIS} camera, to estimate the frequency of fireballs during the PER and GEM activity periods.  

The KoMON software uses an estimate of the apparent angular velocity of a moving object for triggering: as a result, it can capture meteors whose apparent angular velocity (measured for the whole visible meteor trail) is in between 5 and 12 deg/s. Most of the meteors recorded by these cameras are in between 6 and 8 deg/s (note that these are only apparent angular velocities that are not corrected for the elevation of the meteor). As a comparison, the MetRec software uses a minimum angular velocity of 2.5 deg/s for triggering, but it refers to the zenith position, thus, it corresponds to roughly 1-2 deg/s apparent angular velocity. Even though this value is lower than the one used by the KoMON system, the angular velocity distribution of the detected MetRec meteors is quite different: the peak of the distribution is usually around 10 deg/s and the mean value is in between 12 and 14 deg/s. These numbers illustrate that the KoMON cameras are more sensitive to meteors having lower angular velocities. Therefore, by using the term "slow", we mean a meteor that has an apparent angular velocity of about 6-8 deg/s.

Although the KoMON cameras do not provide photometric measurements, we used the same definition for the fireballs as in \cite{deme23}: every meteor that appeared brighter than any of the field stars were considered as a fireball. This is somewhat different from the usual definition (brigher than $-4$ magnitude); the reasons behind the new definition was discussed in \cite{deme23}.
Since there are 5 active KoMON cameras at Piszk\'estet\H{o}, the nightly detection numbers were normalized to a single camera, in order to reduce the statistical bias due to the difference between the FoV of HUPIS and the 5 KoMON cameras that altogether cover the whole sky. By normalizing the fireball numbers to a single KoMON camera, the effective FoV of the two systems becomes about the same. Also, multiple detections, i.e. events captured by more than one camera, were counted as a single one, otherwise they would significantly bias the statistics. 

Figure~\ref{fig:metrec_dslr_comp} shows the normalized KoMON fireball detections as impulses (multiplied by a factor of 100 to enhance their visibility) together with the nightly observed number of meteors detected by {\tt HUPIS}. The meteor shower is indicated at the title of each panel. Note that in 2023 the {\tt HUPIS} camera lost its reference star calibration, which resulted in a significant decrease of the identified PER meteors. This explains the lower peak of the {\tt HUPIS} data in the lower right panel. 

It is clear from Figure~\ref{fig:metrec_dslr_comp} that the fireballs become more frequent during the peak of the shower activity. Assuming that the recorded fireball events were due to shower meteoroids, we estimated the fraction of the fireballs during the showers by taking the ratio of the total number of detected fireballs and the sum of the observed {\tt HUPIS} meteors, after correcting for the different field-of-view of the two system. 
For the PER shower we got  0.37, 0.21 and 0.54 percent for 2021, 2022 and 2023, respectively (again, the 2023 result is overestimated due to the bad performance of {\tt HUPIS} in that year). For the GEM shower we got 0.42 percent in 2021, which is similar to the PER data. These estimates cover the whole activity period, i.e. $\sim 40$ days for the PER and $\sim 10$ days for the GEM shower.

Although the number of detected fireball events are clearly far from being a statistical sample, our data suggest a mean fraction of $\sim 0.3$ percent over $\sim 40$ days for the PER shower and $\sim 0.4$ percent over $\sim 10$ days for the GEM shower meteors. Even though it is difficult to get a realistic estimate for the probability of the impacting events, it might be possible in the future after collecting a significantly larger observational sample.  

\section{Conclusion}

The results of the present study are summarized as follows. 

\begin{enumerate}

\item{We measured the peak ZHR activity of the PER shower between 155 and 185 between 2020 and 2023. These are somewhat higher than other estimates in the literature.
From the measured peak ZHR, the meteoroid fluxes are found in between 31-37 in $10^{-3}$ meteoroid/h/km$^2$ units for meteoroid limiting mass of $2.14 \times 10^{-5}$ g.}

\item{For the GEM shower we were able to obtain useful data only in 2021. We measured the peak ZHR as $221 \pm 9$, which corresponds to a peak meteoroid flux of $\sim 44 \times 10^{-3}$ km$^{-2}$~h$^{-1}$ for meteoroid limiting mass of $1.75 \times 10^{-4}$ g. This is about twice as high as the value reported by \cite{vida22}. The difference may be partly due to our adopted field correction factor ($F = 1.5$), but if we disregard the field correction, our measured values are still somewhat higher. 
}

\item{Using the data from the KoMON meteor camera system, we estimate that the mean fraction of slow fireballs (defined here as slow meteors brighter than 0 magnitude) was  $\sim 0.3$ percent during 40 days for PER and $\sim 0.4$ percent during 10 days for GEM. Even though we could not measure the shower association of the detected fireballs, their increased appearance during the shower maxima suggests that most of them belonged to the active shower. The continuous collection of such data may enable us to estimate or even predict the probability of impacting events in the future.}
\end{enumerate}

Acknowledgements:
We are indebted to an anonymous referee for his/her very helpful comments and suggestions.
This work was supported by the project "Cosmic Effects and Risks" GINOP 2.3.2-15-2016-0003 by the Hungarian National Research Development and Innovation Office, based on funding provided by the European Union. \\
Special thanks are due to Nagykanizsa Amateur Astronomers Association (especially Zsolt Perkó and Attila Gazdag), and the Bárdos Lajos Primary School, Fehérgyarmat (especially Zoltán Pásztor) for their kind contributions.

\bibliography{meteor2-rev.bib}

\begin{thebibliography}{}

\bibitem[{Brown} \& {Rendtel}, 1996]{br96}
{Brown} P. and {Rendtel} J. (1996).
\newblock ``{The Perseid Meteoroid Stream: Characterization of Recent Activity
  from Visual Observations}''.
\newblock {\em Icarus}, {\bf 124:2}, 414--428.

\bibitem[{Ceplecha} et~al., 1998]{ceplecha98}
{Ceplecha} Z., {Borovi{\v{c}}ka} J., {Elford} W.~G., {Revelle} D.~O., {Hawkes}
  R.~L., {Porub{\v{c}}an} V., and {{\v{S}}imek} M. (1998).
\newblock ``{Meteor Phenomena and Bodies}''.
\newblock {\em SSR}, {\bf 84}, 327--471.

\bibitem[{Deme} et~al., 2023]{deme23}
{Deme} L., {Sarneczky} K., {Igaz} A., {Csak} B., {Opitz} N., {Egei} N., and
  {Vinko} J. (2023).
\newblock ``{Comparison of three different camera systems monitoring the meteor
  activity over Hungary in 2020-2023}''.
\newblock {\em WGN, Journal of the International Meteor Organization}, {\bf
  51:6}, 166--174.

\bibitem[{Jenniskens} \& {Miskotte}, 2021]{jm21}
{Jenniskens} P. and {Miskotte} K. (2021).
\newblock ``{Perseid outburst 2021}''.
\newblock {\em eMeteorNews}, {\bf 6:6}, 460--461.

\bibitem[{Koschack} \& {Rendtel}, 1990]{kr90}
{Koschack} R. and {Rendtel} J. (1990).
\newblock ``{Determination of spatial number density and mass index from visual
  meteor observations (II).}''.
\newblock {\em WGN, Journal of the International Meteor Organization}, {\bf
  18:4}, 119--140.

\bibitem[{Miskotte}, 2021]{misk21}
{Miskotte} K. (2021).
\newblock ``{Perseids 2020 revisited}''.
\newblock {\em eMeteorNews}, {\bf 6:1}, 29--30.

\bibitem[{Miskotte} et~al., 2021]{misk21-2}
{Miskotte} K., {Sugimoto} H., and {Martin} P. (2021).
\newblock ``{The big surprise: a late Perseid outburst on August 14, 2021!}''.
\newblock {\em eMeteorNews}, {\bf 6:7}, 517--525.

\bibitem[{Rendtel}, 2004]{rendtel04}
{Rendtel} J. (2004).
\newblock ``{Evolution of the Geminids Observed Over 60 Years}''.
\newblock {\em Earth Moon and Planets}, {\bf 95:1-4}, 27--32.

\bibitem[{Rendtel}, 2023]{rendtel23}
{Rendtel} J. (2023).
\newblock ``{Review of visual meteor observations in 2022}''.
\newblock In {Pajer} U., {Kereszturi} A., {Steyaert} C., {Rendtel} J.,
  {Rudawska} R., {Verbeeck} C., {Gyssens} M., and {Ocana} F., editors, {\em
  Proceedings of the International Meteor Conference, 2022}, pages 130--132.

\bibitem[{Rendtel} \& {Brown}, 1997]{rendtel97}
{Rendtel} J. and {Brown} P. (1997).
\newblock ``{Visual observations of the Perseid meteor shower 1988-1994}''.
\newblock {\em Planetary and Space Science}, {\bf 45:5}, 585--593.

\bibitem[{Vida} et~al., 2022]{vida22}
{Vida} D., {Blaauw Erskine} R.~C., {Brown} P.~G., {Kambulow} J.,
  {Campbell-Brown} M., and {Mazur} M.~J. (2022).
\newblock ``{Computing optical meteor flux using global meteor network data}''.
\newblock {\em MNRAS}, {\bf 515:2}, 2322--2339.

\bibitem[{Watanabe} \& {Marks}, 2018]{watanabe18}
{Watanabe} K.~A. and {Marks} M.~B. (2018).
\newblock ``{Multi-Year Observations of Geminid Meteor Showers with GRT-WF}''.
\newblock {\em WGN, Journal of the International Meteor Organization}, {\bf
  46:5}, 151--153.

\end{thebibliography}



\end{WGNpaper}

\end{document}